\documentclass[aps,prb,twocolumn]{revtex4-1}
\usepackage[english]{babel}
\usepackage[percent]{overpic}

\usepackage{verbatim}
\usepackage[utf8]{inputenc}
\usepackage{amsmath,amssymb,amsfonts}
\usepackage{color}
\usepackage{hyperref}
\usepackage{filecontents}
\usepackage{graphicx}
\usepackage{soul}
\usepackage{bm}
\def\>{\right\rangle}
\def\<{\left\langle}
\def\be{\begin{equation}}
\def\ee{\end{equation}}
\def\ba{\begin{array}{l}}
\def\ea{\end{array}}

\def\beq{\begin{eqnarray}}
\def\eeq{\end{eqnarray}}

\def\iLx{\int dx\hspace{1mm}}

\begin{document}
\title{Spin-thermoelectric transport induced by interactions and spin-flip processes in two dimensional topological insulators}
\author{Flavio Ronetti$^{1,2}$, Luca Vannucci$^{1,2}$, Giacomo Dolcetto$^{2,3}$, Matteo Carrega$^2$, Maura Sassetti $^{1,2}$}
\affiliation{ $^{1} $ Dipartimento di Fisica, Universit\`a di Genova, Via Dodecaneso 33, 16146, Genova, Italy.\\
$^2$ CNR-SPIN, Via Dodecaneso 33, 16146, Genova, Italy.\\
$^3$ Physics and Materials Science Research Unit, University of Luxembourg, L-1511 Luxembourg.} 

\begin{abstract}
	
We consider thermoelectric transport properties of the edge states of a two dimensional topological insulator in a double quantum point contact geometry coupled to two thermally biased reservoirs. Both spin-preserving and spin-flipping tunneling processes between opposite edges are analyzed in the presence of electron-electron interactions. We demonstrate that the simultaneous presence of spin-flipping processes and interactions gives rise to a finite longitudinal spin current. Moreover, its sign and amplitude can be tuned by means of gate voltages with the possibility to generate a pure spin current, with a vanishing charge current. 

\end{abstract}

\maketitle

\section{Introduction}

In the pursuit of high-performance thermoelectric materials a prominent attention has been devoted to topological states of matter, such as quantum Hall systems (QHSs) \cite{Klitzing80,Laughlin81,Halperin82} and topological insulators (TIs) \cite{Dolcetto15,Zhang11,Mele05a,Mele05b,Bernevig06,Konig07}. Thermoelectric effects in QHSs have been investigated for possible future applications exploiting the conduction properties of one dimensional protected chiral edge states \cite{Lopez14,Sanchez15a,Sanchez15b,Hofer15,Vannucci15,carregaetal}. In comparison with QHSs, TIs add to predicted optimal thermoelectric performance \cite{Takahashi10,Takahashi12,Yokoyama14,Xu14} peculiar spin transport properties.
The hallmark of two dimensional (2D) TIs is the emergence of a pair of metallic edge states on the boundaries of the system, where electrons with opposite spin polarization propagate in opposite directions (the so-called spin-momentum locking). Helical edge states, as they are referred to, are topologically protected against backscattering, as long as time-reversal symmetry is preserved. As a consequence, transport along edge states occurs in the ballistic regime and conductance is perfectly quantized. Experimental evidences of 2D TIs were reported through the observation of a quantized conductance in HgTe/CdTe \cite{Konig07,Roth09,Brune12} and InAs/GaSb \cite{Knez11,Knez14,Du15} quantum wells.\\
Spin-momentum locking and topological protection make TIs very interesting materials for the field of spin thermoelectrics \cite{Bauer10,Bauer12}. Thermal manipulation of the spin degree of freedom is relevant to improve spin-based devices, which are believed to be faster and less consuming with respect to their electronic counterparts \cite{Murakami03,Hall06}. Spin thermoelectrics has been recently invigorated by the direct observation of spin Seebeck \cite{Jaworski10,Uchida08}, Peltier \cite{Flipse12} and Seebeck spin tunneling \cite{LeBreton11,Jansen12} effects in magnetic systems. In graphene tunnel junction, spin currents are predicted by coupling to thermally biased normal leads in consequence of intrinsic \cite{Gusynin14} and Rashba \cite{Inglot15} spin-orbit interaction. In HgTe/CdTe quantum wells, a transverse spin current (spin Nerst effect) is expected to be generated in a cross-terminal setup by overlapping of edges state due to finite size effects \cite{Rothe12}.\\
In principle, the linear dispersion relation of edge states in TIs, satisfying  particle-hole symmetry, would prevent any thermoelectric response of edge states. In order to recover thermoelectric effects, previous work proposed to introduce a truncation in the dispersion relation \cite{Takahashi10,Murakami11,Takahashi12}, open a gap \cite{Ghaemi10}, exploit the strong energy-dependence of the lifetime of edge states \cite{Xu14} or introduce a spin-dependent backscattering region through the coupling with an anti-dot \cite{Hwang14}. Also, the combined effects of tunneling and quantum interference in a multiple quantum point contact (QPC) geometry give rise to an energy-dependent transmission amplitude in QHSs, which effectively breaks particle-hole symmetry introducing a thermoelectric response \cite{Vannucci15}. This result is related to topological protection, which guarantees phase-coherent transport regime, allowing for observation of wave-like effects such as quantum interference. Interferometric setups with double QPC junction have been proposed even on the edge states of TIs \cite{Dolcini11,Sternativo13,Dolcini15,Citro11,Romeo12,Ferraro13}. In comparison with QHS interferometry, electrons incoming into a QPC can undergo two distinct tunneling events in TIs. The first is a spin-preserving tunneling process, which induces backscattering events for electrons. The second is a spin-flipping process, which maintains the direction of motion of incoming electrons.  This spin-flipping mechanism arises as a consequence of spin-orbit coupling that can affect the spin quantization axis inside the tunneling region \cite{Teo09,Vayrynen11,Chen16}. The simultaneous presence of quantum interference, which can be properly exploited to break particle-hole symmetry, and tunneling processes related to spin polarization makes interferometric setup in TIs appealing for spin thermoelectrics.\\
In this paper, we investigate spin thermoelectrics in a 2D TI connected to two thermally biased reservoirs in the presence of a double QPC geometry.  Our aim is to put the focus on the role of electron-electron interactions in spin transport properties of helical edge states. This purpose is strongly supported by the recent experimental observation of interaction-induced effects on the conductance of the edge states in InAs/GaSb heterostructures, where power law behaviors typical of a Luttinger liquid were measured at low temperatures \cite{Li15}.\\ Remarkably, we found that interactions play a crucial role in the generation of a spin thermoelectrics response in TIs. We demonstrate that the contribution of spin-flipping tunneling to spin transport in a two terminal configuration is dramatically modified by the introduction of interactions. The effective transmission related to spin-flipping processes acquires a functional dependence on energy in the interacting regime. This energy dependence, necessary in order to generate thermoelectric effects, cannot be observed in non-interacting systems \cite{Sternativo13}. In consequence of this novel property, a longitudinal thermoelectric spin current is generated in response to a thermal gradient in the non linear regime through spin-flipping tunneling.\\
Finally, we show that pure spin currents (i.e.~non-vanishing spin currents in the absence of any charge current) can be generated in our setup by properly acting on the parameters of the tunneling region. \\
The paper is organized as follows. In Sec. \ref{sec:setup} we introduce the setup and the tunneling operators for spin-preserving and spin-flipping processes. In Sec. \ref{sec:c/s_currents} we define current operators and discuss transmission amplitudes in the non interacting case. In Sec. \ref{sec:interactions} we focus on the properties induced by interactions and describe charge and spin transport in the framework of helical Luttinger liquid model. Sec. \ref{sec:results} is devoted to the evaluation of thermoelectric currents and to the presentation of the results for a double QPC geometry. Finally, in Sec. \ref{sec:conclusions}, we draw our conclusion.

\section{Setup and general settings}
\label{sec:setup}

\begin{figure}[h]
\centering
\includegraphics[width=.9\linewidth]{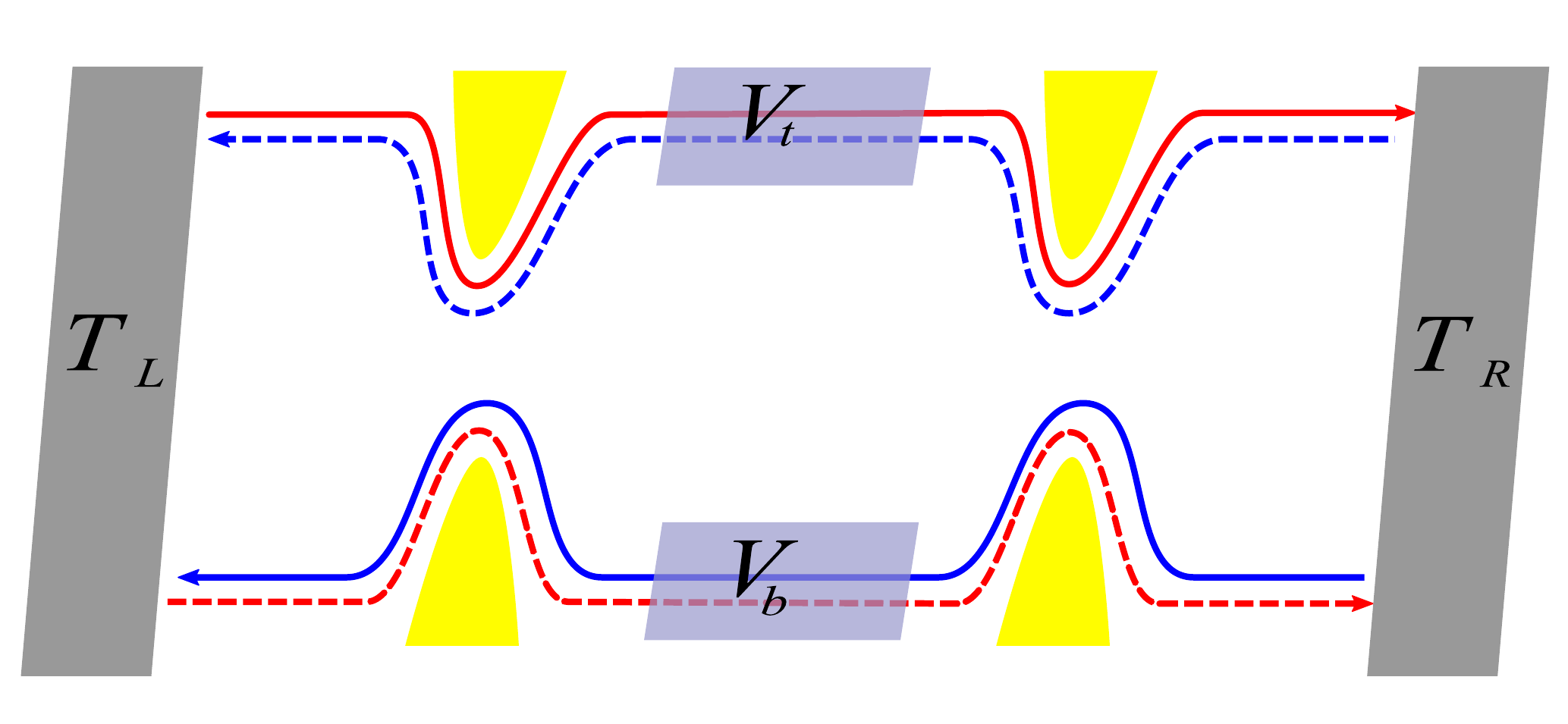} 
\caption{{(Color online) Scheme of the setup. A 2D TI in a two terminal configuration with fixed chemical potential $\mu_R=\mu_L=\mu$ and a finite thermal gradient $T_L \neq T_R$ is considered. On each edge, electrons with spin $\uparrow$ (solid line) and spin $\downarrow$ (dashed line) counterpropagate. Two quantum point contacts pinch the two edges allowing for electron tunneling events. Gate voltages $V_t$ and $V_b$ locally modify electron density in
the region between the two QPCs .}}
\label{fig:one}
\end{figure}

We consider a 2D TI in a two terminal configuration, as depicted in Fig.~\ref{fig:one}. The two contacts have the same chemical potential $\mu_L=\mu_R=\mu$ but two different temperatures $T_L$ and $T_R$, with $L$ ($R$) indicating the left (right) reservoir.
Here the direction of propagation of spin $\uparrow$ and spin $\downarrow$ electrons, in both edges, is constrained by helicity. The free Hamiltonians of each channel $(r,\sigma$), with $r=R,L$ and $\sigma=\uparrow,\downarrow$ are ($\xi_{R/L}=\pm 1$ and $\hbar=1$)
\begin{equation}
\label{freeham}
H_{r,\sigma}=v_{\rm F}\iLx \psi^{\dagger}_{r,\sigma}(x)\left(-i \xi_r \partial_x\right)\psi_{r,\sigma}(x),
\end{equation}
where $v_{\rm F}$ is the Fermi velocity and $\psi_{r,\sigma}(x)$ is the fermionic field which annihilates right-moving ($R$) or left-moving ($L$) electrons with spin $\uparrow$ or $\downarrow$.
The Hamiltonian Eq.~\eqref{freeham} conserves the number operator
\begin{equation}
\label{opnum}
N_{r,\sigma}=\iLx \rho_{r,\sigma}(x) = \iLx : \psi^{\dagger}_{r,\sigma}(x)\psi_{r,\sigma}(x):~,
\end{equation}
which counts the number of electrons moving in the $r$ direction with spin $\sigma$. The presence of constrictions, like quantum point contacts, allows for interedge tunneling events. Time-reversal symmetry restricts the possible tunneling events to \cite{Citro11,Ferraro13}
\begin{equation}
\label{pham}
H_p=\Lambda_{p}\sum_{\sigma=\uparrow,\downarrow}\int dx\hspace{1mm}f(x)\psi_{L,\sigma}^{\dagger}(x)\psi_{R,\sigma}(x)+\text{h}.\text{c.},
\end{equation}
and
\begin{equation}
\label{fham}
H_f=\Lambda_{f}\sum_{r=R,L}\xi_{r}\int dx\hspace{1mm}f(x)\psi_{r,\uparrow}^{\dagger}(x)\psi_{r,\downarrow}(x)+\text{h}.\text{c.}.
\end{equation}
The former accounts for spin-preserving ($p$) backward scattering processes, while the latter takes into account spin-flipping ($f$) forward scattering. In the above equations $\Lambda_{p/f}$ are the constant tunneling amplitude for $p/f$ processes, whereas $f(x)$ describes the precise shape of the tunneling region~\cite{Vannucci15,Dolcetto12}. 
Note that Eqs.~\eqref{pham} and \eqref{fham} conserve the total number of electrons on each edge separately and satisfy
\begin{align}
\label{SPrelation}&\dot{N}_{R,\sigma}+\dot{N}_{L,\sigma}=0 \hspace{5mm} \sigma=\uparrow,\downarrow,\\\label{SFrelation}
&\dot{N}_{r,\uparrow}+\dot{N}_{r,\downarrow}=0 \hspace{5mm} r=R,L, .
\end{align}
Although the formalism is valid for a generic tunneling region, we will present results  for the case of a double QPC geometry with $f(x)=\frac{1}{2}\sum\limits_{p=\pm1}\delta(x-p\frac{d}{2})$, where $d$ is the distance between the two QPCs.
{As previously proposed\cite{Dolcini11,Virtanen11,Ferraro13,Sternativo13}, we consider two gate voltages $V_{t/b}$ capacitively coupled to the top and bottom edges to shift the Fermi momenta
\begin{equation}
k_{\rm F}^{(t/b)}=k_{\rm F}+\frac{e V_{t/b}}{v_{\rm F}},
\end{equation}
thus effectively modifying the dynamical phase acquired by the propagating electrons.}

\section{Charge and spin currents}
\label{sec:c/s_currents}

We are interested in studying whether a thermal gradient $T_L\neq T_R$ is able to induce charge $I_c$ and spin $I_s$ currents flowing along the edges of the system.
In the absence of tunneling, spin current $I_s$ is zero in a two terminal configuration due to symmetry constraints. The same occurs for charge current $I_c$, because particle-hole symmetry is preserved and prevents any thermoelectric effect due to a thermal gradient. In the following we investigate the generation of thermoelectric currents in the presence of interedge tunneling.
We thus define charge and spin current operators as ($e>0$)
\begin{align}
\label{chargecurrent}
I_c&\equiv -\frac{e}{2}\left(\dot{N}_{R,\uparrow}+\dot{N}_{R,\downarrow}-\dot{N}_{L,\uparrow}-\dot{N}_{L,\downarrow}\right),\\
\label{spincurrent}
I_s&\equiv \frac{1}{4}\left(\dot{N}_{R,\uparrow}+\dot{N}_{L,\downarrow}-\dot{N}_{L,\uparrow}-\dot{N}_{R,\downarrow}\right),
\end{align}
which, by virtue of particle number conservation, can be recast in the form
\begin{align}
\label{chargecurrent2}
I_c&= -e\left(\dot{N}_{R,\uparrow}+\dot{N}_{R,\downarrow}\right),\\
\label{spincurrent2}
I_s&= \frac{1}{2}\left( \dot{N}_{R,\uparrow}+\dot{N}_{L,\downarrow}\right).
\end{align}
Since we are interested in evaluating the expectation value of Eqs.~\eqref{chargecurrent2}-\eqref{spincurrent2}, it is useful to express them in terms of fermionic operators
\begin{align}
I_c=& -ie \Lambda_p\sum\limits_{\sigma}\int dx \hspace{1mm} f(x)  \psi_{L,\sigma}^{\dagger}(x)\psi_{R,\sigma}(x)+\text{h}.\text{c.},\\
I_s=&\frac{i}{2}\Bigg(\Lambda_{p}\sum\limits_{\sigma}\xi_{\sigma}\int dx \hspace{0mm}f(x)\psi^{\dagger}_{L,\sigma}(x)\psi_{R,\sigma}(x)+\nonumber\\&-\Lambda_{f}\sum\limits_{r}\int dx \hspace{0mm}f(x) {\psi^{\dagger}_{r,\uparrow}}(x)\psi_{r,\downarrow}(x)\Bigg)+\text{h}.\text{c}.
\end{align}
It is worth noting that charge current is unaffected by $f$-tunneling processes.

\subsection{Transmission amplitudes in the non-interacting case}

The transport properties of the system crucially depend on the presence of tunneling. Therefore it is useful to recall the physical picture for the $p$ and $f$ processes, focusing, for the moment, on the case without electron interactions~\cite{Sternativo13}, whose presence will be discussed later. In the absence of interactions the scattering matrix formalism can be applied, allowing to evaluate the transmission through the tunneling regions. It is possible to demonstrate that, for an arbitrary tunneling region, the transmission amplitude factorizes into two contributions ${\cal T}_p$ and ${\cal T}_f$, related to spin-preserving and spin-flipping processes respectively \cite{Sternativo13}. We briefly recall the main results and the physical mechanism for these two quantities. We restrict the discussion to weak tunneling between the edges, which will allow a comparison with the interacting regime discussed in the next Section.
\begin{figure}[h]
\centering
\includegraphics[scale=0.25]{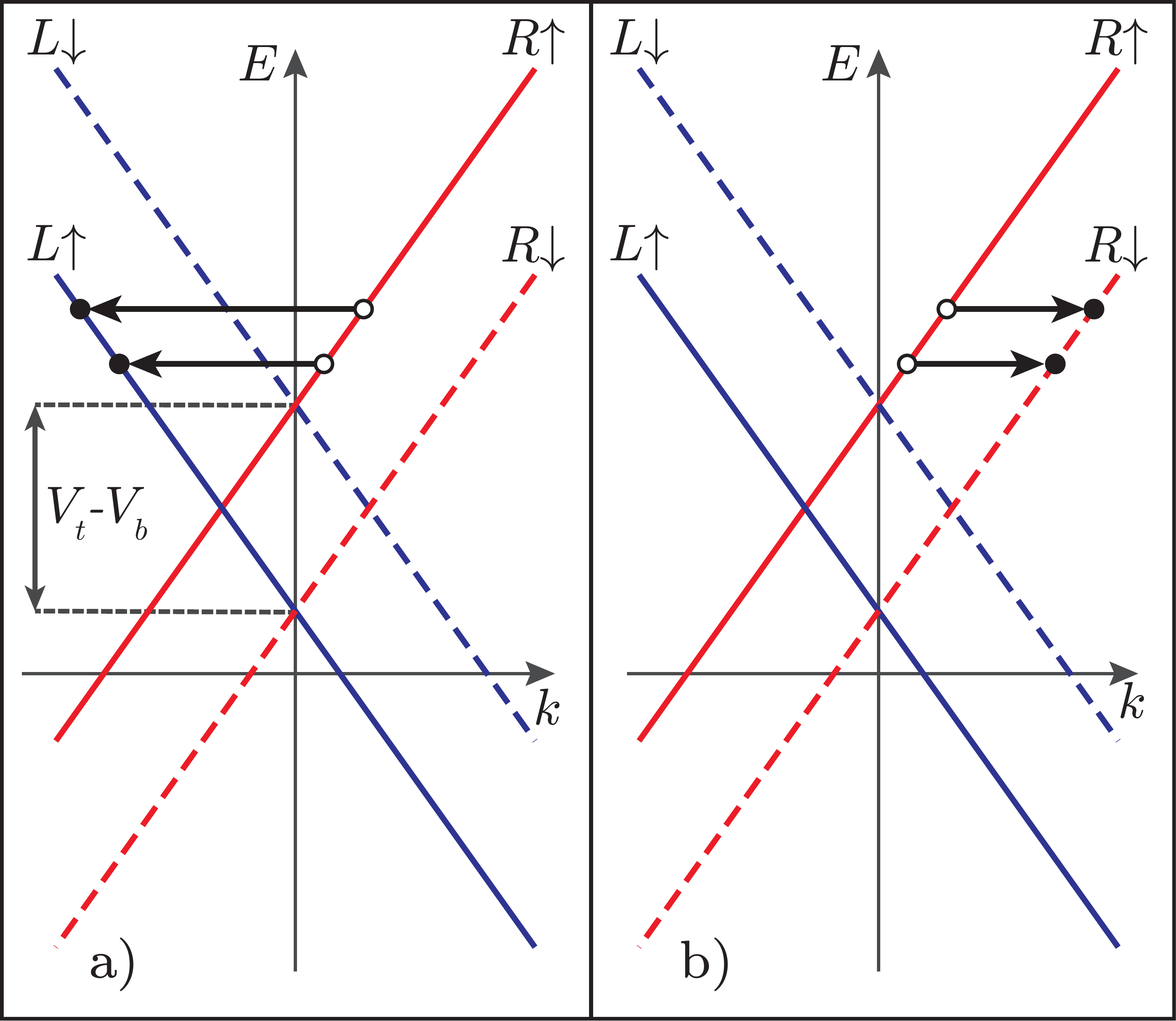}
\caption{(Color online) Dispersion relation of edge states shifted by gate voltages inside the tunneling region. (a) $p$-tunneling: two $R\uparrow$ electrons move to $L\uparrow$ branch, changing their chirality; the transferred momentum depends on energy. (b) $f$-tunneling: two $R\uparrow$ electrons move to $R\downarrow$ branch, changing their spin; the transferred momentum does not depend on energy.}
\label{fig:two}
\end{figure}In Fig.~\ref{fig:two} is depicted the energy dispersion for edge states in the tunneling region. The presence of gate voltages shifts the Dirac cones. Let us consider firstly $p$-tunneling, depicted in Fig.~\ref{fig:two}(a). In this case a right moving excitation can tunnel to the other edge in the opposite branch preserving its spin component. Contrary to energy, momentum is not conserved during tunneling because of the breaking of translational invariance. Assuming that an electron with energy $E$ tunnels from the top to the bottom edge, the transferred momentum will be
\begin{equation}
q=k'-k''= 2\frac{E}{v_{{\rm F}}} + k_{{\rm F}}^{(t)} + k_{{\rm F}}^{(b)}~,
\end{equation}
and it depends on the energy $E$.\\ Nevertheless the functional dependence of the transmission amplitude ${\cal T}_p(E)$ is crucially determined by the shape of the tunneling region. Indeed, the transferred momentum appears exclusively in the electronic phases, which affect transmission amplitude solely in the presence of quantum interference effects. Specific tunnel junction, such as multiple QPC or extended contact junctions, are required to observe quantum interference. Under these conditions, $p$-tunneling occurs with an energy-dependent transmission amplitude ${\cal T}_p(E)$.\\
In contrast, $f$-tunneling processes cannot give rise to any energy-dependent transmission amplitude ${\cal T}_f$. In Fig.~\ref{fig:two}(b) we show such processes, where electrons tunneling between the two edges change their spin components but not their chiralities. Looking at Fig.~\ref{fig:two}(b) one can argue that energy conserving processes all transfer the same momentum, and therefore transmission amplitude ${\cal T}_f$ is not energy dependent.\\
Consider now the effects of $p$- and $f$-tunneling processes on the charge and spin currents. A $R \uparrow$ excitation can reach the right reservoir either remaining on its channel or by tunneling via $f$-process, changing its spin component into $R \downarrow$. The former situation is described by the transmission amplitude ${\cal T}_p {\cal T}_f$, while the latter reads ${\cal T}_p (1-{\cal T}_f)$. Therefore the total transmission for a $R\uparrow$ electron from left to right reservoir is given by ${\cal T}={\cal T}_p {\cal T}_f + {\cal T}_p(1-{\cal T}_f)={\cal T}_p$, and it is {\it not} affected by $f$-processes. According to Landauer-B\"uttiker formalism \cite{Buttiker}, the corresponding expectation value can be written as
\begin{align}
\langle\dot{N}_{R\uparrow}\rangle&=\int dE \hspace{1mm}\mathcal{T}\left[n_L(E)-n_R(E)\right]=\nonumber\\\label{nonintRup}&=\int dE \hspace{1mm}\mathcal{T}_p(E)\left[n_L(E)-n_R(E)\right],
\end{align}
which is different from zero, since ${\cal T}_p(E)$ transmission for $p$-processes is energy dependent and breaks particle-hole symmetry. A similar argument holds true also for $R \downarrow$ and $L \downarrow$ electrons, whose expectation values are $\langle\dot{N}_{R\downarrow}\rangle=-\langle\dot{N}_{L\downarrow}\rangle=\langle\dot{N}_{R\uparrow}\rangle$.
By means of definitions \eqref{chargecurrent2} and \eqref{spincurrent2} we see that longitudinal spin current $I_s$ is zero, while charge current can be written as
\begin{equation}
\label{landauercharge}
\langle I_c\rangle=-2e\int dE \hspace{1mm}\mathcal{T}_p(E)\left[n_L(E)-n_R(E)\right].
\end{equation}
Although breaking of particle-hole symmetry due to tunneling is able to generate charge thermoelectric effects, the spin sector remains insensitive to the thermal gradient because of symmetry considerations.\\ In the following we will show that electron-electron interactions can radically modify this picture.

\section{Turning on interactions}
\label{sec:interactions}

Recently, evidence of electron interactions at the edge of the 2D TI InAs/GaSb has been reported~\cite{Li15}. Therefore it is important to investigate how their presence affect the thermoelectric and transport properties discussed in the previous section.
The Hamiltonian for the interacting electrons is thus
\begin{equation}
H=\hspace{-2mm}\sum_{\substack{r=R,L\\\sigma=\uparrow,\downarrow}}\hspace{-2mm}H_{r,\sigma} + H_{{\rm int}}~,
\end{equation}
with $H_{r,\sigma}$ the free Hamiltonians in Eq.~\eqref{freeham} and
\beq
&& H_{{\rm int}}= \frac{g_4}{2}\sum_{r,\sigma}\int dx \left(\rho_{r,\sigma}(x)\right)^2 + \nonumber \\
&+& g_2 \sum_\sigma \int dx \rho_{R,\sigma}(x) \rho_{L,-\sigma}(x)~,
\eeq
where $g_4$ and $g_2$ are coupling constants describing the strength of intra- and inter-channel interactions among electrons. Interactions are treated in the so-called helical Luttinger liquid model, and the above Hamiltonian can be diagonalized introducing proper bosonic fields and using well-known bosonization techniques \cite{Vondelft,Voit95,Calzona15}. Here, fermionic operators can be expressed as
\begin{subequations}
\begin{align}
\psi_{R\uparrow}(x)&=\frac{\mathcal{F}_{R,\uparrow}}{\sqrt{2\pi a}}e^{ik_{{\rm F}}^{(t)}x}e^{-i\sqrt{2\pi}\phi_{R,\uparrow}(x)},\\
\psi_{L\uparrow}(x)&=\frac{\mathcal{F}_{L,\uparrow}}{\sqrt{2\pi a}}e^{-ik_{{\rm F}}^{(b)}x}e^{-i\sqrt{2\pi}\phi_{L,\uparrow}(x)},\\
\psi_{R\downarrow}(x)&=\frac{\mathcal{F}_{R,\downarrow}}{\sqrt{2\pi a}}e^{ik_{{\rm F}}^{(b)}x}e^{-i\sqrt{2\pi}\phi_{R,\downarrow}(x)},\\
\psi_{L\downarrow}(x)&=\frac{\mathcal{F}_{L,\downarrow}}{\sqrt{2\pi a}}e^{-ik_{{\rm F}}^{(t)}x}e^{-i\sqrt{2\pi}\phi_{L,\downarrow}(x)},
\end{align}
\end{subequations}
with $\phi_{r,\sigma}$ bosonic fields, $\mathcal{F}_{r,\sigma}$ the so-called Klein factors and $a$ a short-length cut-off.
The new chiral fields, satisfying $\phi_{\pm}^{(t/b)}(x,t)=\phi_{\pm}^{(t/b)}(x \mp ut)$, are \cite{Miranda03,Voit95}
\begin{align}
&\phi^{(t)}_{\pm}(x)= \frac{1\pm\mathcal{K}}{2\sqrt{\mathcal{K}}}\hspace{0.75mm}\phi_{R,\uparrow}(x)+\frac{1\mp\mathcal{K}}{2\sqrt{\mathcal{K}}}\hspace{0.75mm}\phi_{L,\downarrow}(x),\\
&\phi^{(b)}_{\pm}(x)= \frac{1\pm\mathcal{K}}{2\sqrt{\mathcal{K}}}\hspace{0.75mm}\phi_{R,\downarrow}(x)+\frac{1\mp\mathcal{K}}{2\sqrt{\mathcal{K}}}\hspace{0.75mm}\phi_{L,\uparrow}(x),
\end{align}
where $u = v_{\rm F} \sqrt{\left(1+\frac{g_4}{2\pi v_{\rm F}}\right)^2 - \left(\frac{g_2}{2\pi v_{\rm F}}\right)^2}$ is the renormalized velocity and
\begin{equation}
\label{luttinger}
\mathcal{K}=\sqrt{\frac{2\pi v_{\rm F} + g_4-g_2}{2\pi v_{\rm F} +g_4+g_2}}\leq 1~,
\end{equation}
is the interaction parameter. In terms of chiral bosonic fields the uncoupled interacting edge states are described by
\begin{equation}
\label{hamdiag}
H=\sum_{\ell=t,b}\frac{u}{4\pi}\int dx \hspace{1mm}\left[\left(\partial_x\phi_{+}^{(\ell)}(x)\right)^2+\left(\partial_x\phi_{-}^{(\ell)}(x)\right)^2\right]
.\end{equation}
{In the following analysis we study the weak tunneling regime by considering not too strong interactions, namely $1/\sqrt{3}\leq {\cal K}\leq 1$.
In this regime, single electron tunneling is in general\cite{note} the dominant contribution, both in the single and in the double constriction setup\cite{Mirlin13,kane92,Braggio12,Furusaki93,Sassetti95}.
Multi-particle processes can also occur\cite{Teo09,Schmidt11,Mirlin13}, but at low energies and weak interactions they are expected to be negligible compared to single-particle tunneling.}\\
Let us discuss now how $f$-tunneling is affected by the presence of interactions. Recalling the simple scheme depicted in Fig.~\ref{fig:two}(b), in the non-interacting case (${\cal K}=1$) the final energy of the $R\downarrow$ electron is fixed and correspond to the incoming one of $R\uparrow$ electron. In that case, processes in which the $R\downarrow $ electron has lower energy than the $R\uparrow $ one are not allowed. On the contrary, in the interacting case (${\cal K}<1$) such processes exist, due to the peculiar broadening of the energy spectrum of helical Luttinger liquids \cite{Imambekov12, Voit93}. For these processes a $R\uparrow$ electron tunnels into the $R\downarrow$ channel, losing a certain amount of energy and creating a particle-hole excitation in the $L\uparrow$ channel. The missing energy is stored into the particle-hole contribution in the opposite channel. The creation of such particle-hole excitations is due to the peculiar form of the interaction, and in particular to the term of inter-channel density-density interactions,  proportional to $g_2$ \cite{Voit93}.\\
Each tunneling event of the type described above is associated with a finite energy-dependent transferred momentum. Therefore, because of the spectral function broadening of the linear dispersing edge states in the presence of interactions, $f$-events can depend on the energy at which tunneling occurs.\\
Furthermore, let us notice that tunneling of a $R\uparrow$ electron induces a finite transferred momentum on each branch of the bottom edge, in the presence of interactions. This means that a single $f$-tunneling event induces both a forward and a backward scattering, in contrast to the same process in the non-interacting case. As a consequence, it is no longer true that charge and spin are transferred from one reservoir to the other independently of $f$-tunneling.\\ From these considerations,  we predict that electron interactions dramatically affect the thermoelectric charge and spin response.
In particular, we will show that the interactions-induced energy dependence of $f$-processes can lead to a thermally driven spin current.

\section{Evaluation of thermoelectric spin current}
\label{sec:results}

In this section we calculate average currents in the setup of Fig.~\ref{fig:one} in presence of electron-electron interactions. We evaluate the expectation value of current operators \eqref{chargecurrent2} and \eqref{spincurrent2} at lowest order in the tunneling \cite{Kamenev09,Martin-notes}. We obtain
\begin{align}
\langle I_c\rangle=& -4ei\left|\lambda_p\right|^2\int \hspace{0mm}dx\int \hspace{0mm}dx' f(x) f(x')\nonumber\\
&\times\int dt\hspace{1mm}\sin\left[\left(k_{{\rm F}}^{(t)}+k_{{\rm F}}^{(b)}\right)(x-x')\right]\nonumber\\
&\times P_{+,\gamma}\left(t-\frac{x-x'}{u}\right)P_{-,\gamma}\left(t+\frac{x-x'}{u}\right),
\end{align}
for charge current, and
\begin{align}
\langle I_s\rangle&=i\left|\lambda_f\right|^2\int dx \int dx' f(x) f(x')\nonumber\\
&\times\int dt \hspace{1mm}\sin\left[\left(k_{{\rm F}}^{(t)}-k_{{\rm F}}^{(b)}\right)\left(x-x'\right)\right] \nonumber\\
&\times \Bigg\{P_{+,\gamma_1} \left(t-\frac{x-x'}{u}\right) P_{-,\gamma_2}\left(t+\frac{x-x'}{u}\right)+\nonumber\\
&+P_{-,\gamma_1}\left(t-\frac{x-x'}{u}\right)P_{+,\gamma_2}\left(t+\frac{x-x'}{u}\right)\Bigg\},
\end{align}
for spin current, where $\lambda_{p/f}=\frac{\Lambda_{p/f}}{2\pi a}$ and we have introduced 
\be
\gamma=\frac{\gamma_1 + \gamma_2}{2} \qquad \gamma_{1/2} = \frac{1}{2}\big({\cal K} + \frac{1}{{\cal K}} \pm 2\big)~.
\ee
The contribution of $p$-tunneling to average spin current is zero due to the symmetry between $\uparrow$ and $\downarrow$ electrons in a two terminal configuration.\\ The function $P_{\pm,g}(t)=e^{g \mathcal W_\pm (t)}$ is related to the bosonic correlation function \cite{weiss-book,giamarchi-book,Kleimann02} ($k_{\rm B} = 1$)
\begin{equation}
\mathcal W_{\pm}(t)=\ln{\frac{\left|\Gamma\left(1+\frac{T_{L/R}}{\omega_c}+iT_{L/R} t\right)\right|^2}{\Gamma^2\left(1+\frac{T_{L/R}}{\omega_c}\right)\left(1+i\omega_c t\right)}},
\end{equation}
with $\mathcal W_\pm (t) = \< \left[\phi_\pm^{(\ell)} (t) - \phi_\pm^{(\ell)} (0)\right] \phi_\pm^{(\ell)} (0) \> $ equal for the top and bottom edges and $\omega_c=u/a$ the high-energy cut-off. It is convenient to switch to the energy representation, by using the Fourier transform $\tilde{P}_{\pm , g}(E)=\int dt e^{-i E t}P_{\pm , g}(t)$. In the limit $\omega_c/T_{L/R}\gg 1$, it can be written as
\begin{equation}
\label{fermisub}
\tilde{P}_{\pm,g}(E)=\mathcal{D}_{\pm,g}(E)n_{L/R}(E),
\end{equation}
where $n_{L/R}(E)$ is the Fermi distribution~\cite{footnote} of left and right reservoirs and
\begin{align}
\mathcal{D}_{\pm,g}(E)&=2\cosh\left[\frac{E}{2T_{L/R}}\right]\left(\frac{2\pi T_{L/R}}{\omega_c}\right)^g\frac{1}{2\pi T_{L/R}} \nonumber\\
&\times \mathcal{B}\left[\frac{g}{2}-i\frac{E}{2\pi T_{L/R}},\frac{g}{2}+i\frac{E}{2\pi T_{L/R}}\right] ~,
\label{de}
\end{align}
with $\mathcal{B}(x,y)$ the Euler Beta function.
The above function plays the role of an effective density of state for an interacting system, with energy and temperature dependence, and it reduces to a constant in the limit $g\to 1$.
In the energy representation the average currents read
\begin{align}
\label{corrente1pspcarica}
\langle I_{c} \rangle= & \int \frac{dE}{2\pi}\hspace{1mm} g_p(E)\mathcal{D}_{+,\gamma}(E)\mathcal{D}_{-,\gamma}(E)\left[n_L(E)-n_R(E)\right],\\
\label{corrente1psfspin}
\langle I_s \rangle = & \int\frac{dE}{2\pi}\hspace{1mm}g_f(E)\left[ \mathcal{D}_{+,\gamma_1}(E)\mathcal{D}_{-, \gamma_2}(E) \right. \nonumber\\
& \left.-\mathcal{D}_{+,\gamma_2}(E)\mathcal{D}_{-,\gamma_1}(E)\right] \left[n_L(E)-n_R(E)\right].
\end{align}
The functions
\begin{align}
	\label{eq:non-int}
	\begin{pmatrix} g_p(E) \\ g_f(E) \end{pmatrix} &=  \int dx\hspace{1mm} \int  dx'\hspace{1mm} f(x) f(x')\sin\left[\frac{2E}{u}(x-x')\right]\nonumber\\
	&\times\begin{pmatrix} -2e\left|\lambda_p\right|^2 \sin\left[\left(k_{\rm F}^{(t)}+k_{\rm F}^{(b)}\right)(x-x')\right] \\ \frac{\left|\lambda_f\right|^2}{2} \sin\left[\left(k_{\rm F}^{(t)}-k_{\rm F}^{(b)}\right)(x-x')\right] \end{pmatrix}.
\end{align}
play the role of effective transmission probabilities for charge and spin transport respectively. Their energy dependence can be modified by changing the shape of the tunneling region, and this is crucial in order to allow for thermoelectric effects. Both charge and spin currents acquire a power-law dependence with temperature determined by the functions $\mathcal{D}_{\pm ,g }(E)$, with the exponent linked to the interaction strength, as typical of Luttinger liquids.\\ Looking at Eqs.~\eqref{corrente1pspcarica} and \eqref{corrente1psfspin}, one immediately sees that the charge current is an odd function of the temperature difference $\Delta T = T_R - T_L$, while the spin current is an even function of $\Delta T$. This means that the thermoelectric generation of a spin current requires necessarily a non-linear regime.\\
It is worth to note that, putting $\mathcal{K}=1$ in Eq. \eqref{corrente1pspcarica}, we can identify by comparison with Eq. \eqref{landauercharge} the $p$-transmission amplitude at the lowest order in tunneling \cite{Sternativo13}
{
\begin{align}
& \mathcal{T}_p(E)=\frac{2\pi \left|\lambda_p\right|^2}{\omega_c^2} \int dx \int dx' f(x) f(x')\nonumber\\& \times \sin\Big[\Big(k_{\rm F}^{(t)}+k_{\rm F}^{(b)}\Big)(x-x')\Big]   \sin\left[\frac{2E}{v_{\rm F}}\left(x-x'\right)\right].
\end{align}
}

\subsection{Double quantum point contact geometry}

\begin{figure}
\centering
\includegraphics[width=.85\linewidth]{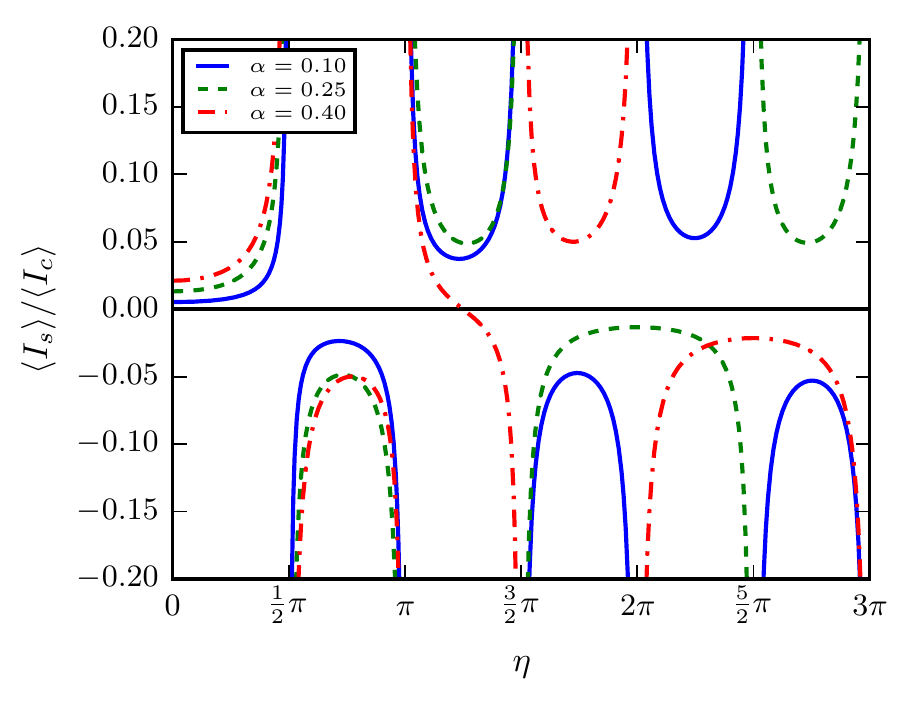} 
\caption{(Color online) The plot shows the ratio $\langle I_s \rangle/\langle I_c \rangle$ as a function of $\eta$, in units of $|\lambda_f|^2/(4e|\lambda_p|^2)$. Different curves refer to different values of $\alpha=0.1;0.25;0.4$. Temperatures are set to $T_L=0.1$~K and $T_R=1$~K, while the interaction strength is ${\cal K}=0.6$. The chemical potential and the high energy cut-off are $\mu=100$ K and $\omega_c=200$ K.
}
\label{fig:ratio}
\end{figure}

\begin{figure}
\centering
\includegraphics[width=.85\linewidth]{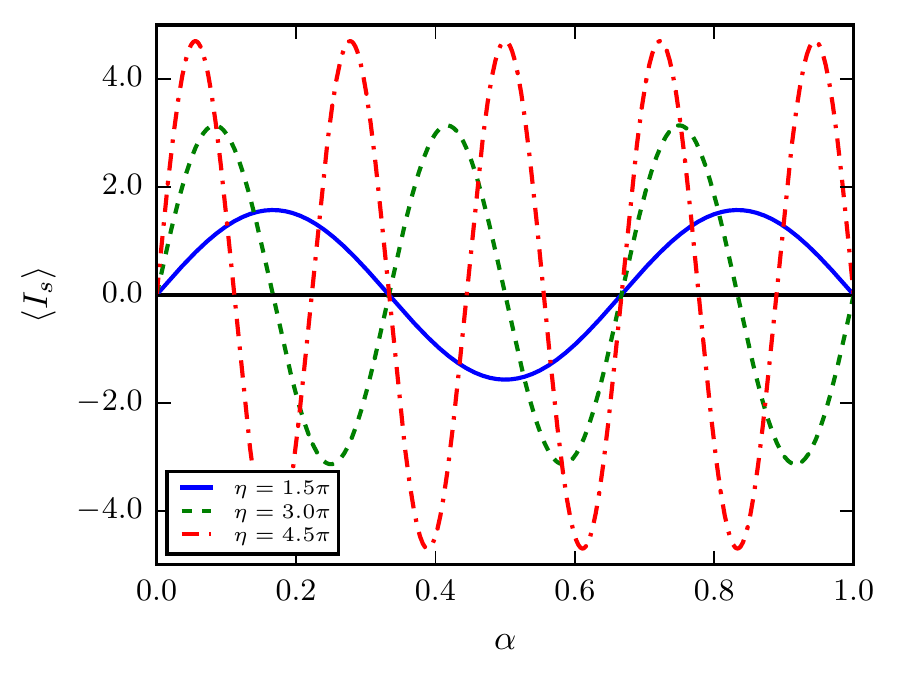}
\caption{(Color online) Average spin current $\langle I_s \rangle$ in units of $10^{-6} |\lambda_f|^2/(8\pi)$ as a function of the parameter $\alpha$. Different curves refer to different values of $\eta=3\pi/2 ; 3\pi ; 9\pi/2$. In all cases $\langle I_c \rangle=0$ with a finite pure spin current. Other parameters are the same as in Fig.~\ref{fig:ratio}. 
}
\label{fig:alpha}
\end{figure}

To make quantitative prediction, we now focus on tunneling region composed by two QPCs placed at position $x=\pm d/2$.
Before starting, we recall that for a single QPC $f(x)=\delta(x)$, and both currents $\langle I_{c/s}\rangle $ vanish,  since the transmission functions are energy independent and can't give rise to any thermoelectric effect.
\begin{figure}[h]
\centering
\includegraphics[width=0.9\linewidth]{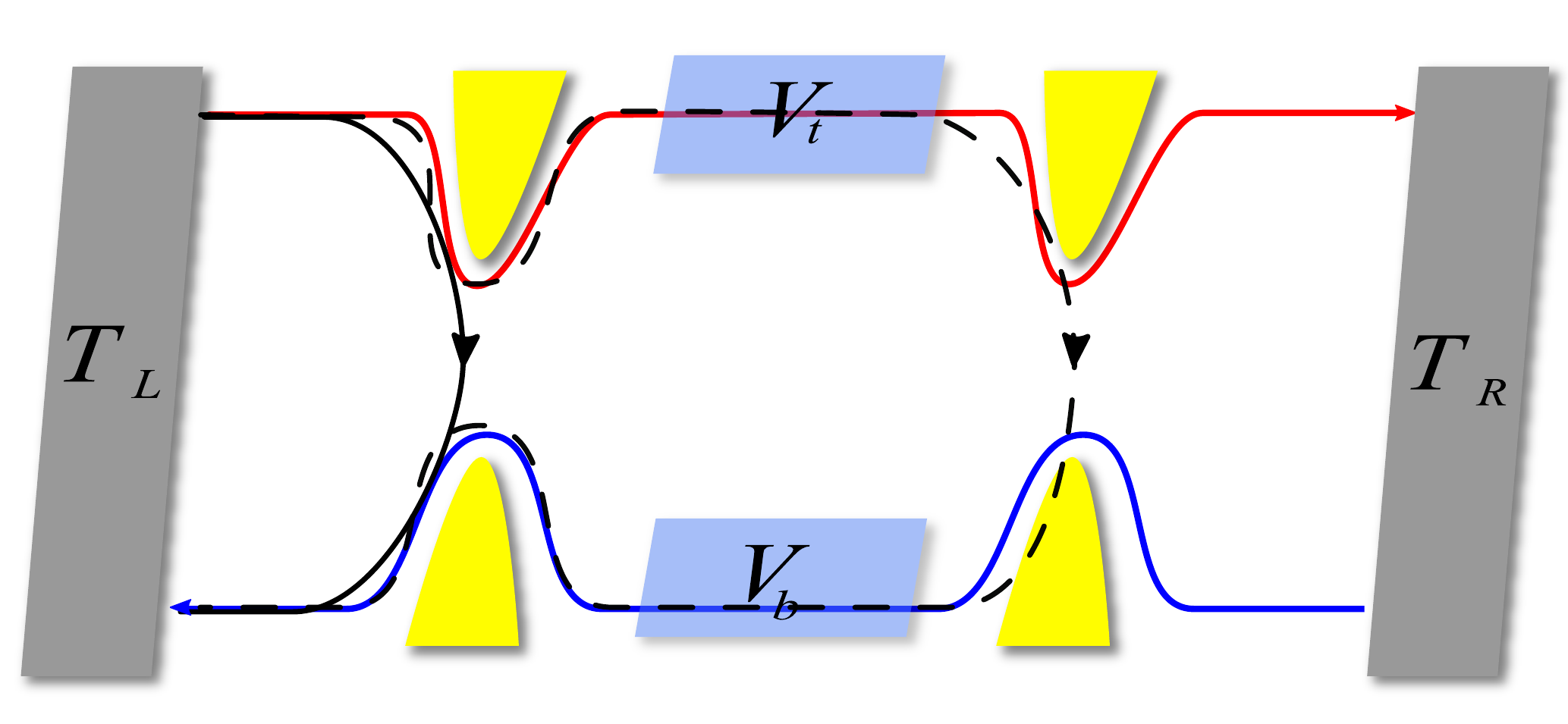} 
\caption{{(Color online) Two possible paths for a $p$-tunneling event. A $R\uparrow$ electron (solid red line) can tunnel to $L\uparrow$ (solid blue line) through left (black solid line) or right (black dashed line) QPC.}} 
\label{2QPC}
\end{figure}In the case of interest with two QPCs, the function $f(x)=\frac{1}{2}\sum\limits_{p=\pm1}\delta\left(x-p~d/2\right)$ and thus~\cite{footnote2}
\begin{align}
\< I_c \> & = -\frac{e|\lambda_p|^2}{2\pi} \sin(2\eta) \int dE \sin \left(2\eta \frac{\mathcal K E}{\mu}\right) \nonumber \\
& \times \mathcal D_{+,\gamma}(E) \mathcal D_{-,\gamma}(E) [n_L(E) - n_R(E)],\\ \nonumber\\
\label{I_spin_2qpc}
\< I_s \> & = \frac{|\lambda_f|^2}{8\pi} \sin(2\alpha \eta) \int dE \sin \left(2\eta \frac{\mathcal K E}{\mu}\right) \nonumber \\
& \times \left[ \mathcal D_{+,\gamma_1}(E) \mathcal D_{-,\gamma_2}(E) - \mathcal D_{+,\gamma_2}(E) \mathcal D_{-,\gamma_1}(E) \right] \nonumber \\
& \times [n_L(E) - n_R(E)].
\end{align}
The dimensionless parameters
\be
\eta=\frac{k_{{\rm F}}^{(t)}+k_{{\rm F}}^{(b)}}{2}d , \qquad \qquad \alpha =\frac{k_{{\rm F}}^{(t)}-k_{{\rm F}}^{(b)}}{k_{{\rm F}}^{(t)}+k_{{\rm F}}^{(b)}} ,
\ee
have been defined, which are related to the distance between the QPCs and to the gate voltage configuration.\\
The region between the two QPCs produces interference pathways for electrons, as schematically depicted in Fig. \ref{2QPC}. 
A $R\uparrow$ electron starting from left reservoir can undergo a $p$-event either at the left (black solid line in Fig. \ref{2QPC}) or at the right (black dashed line in Fig. \ref{2QPC}) QPC. The energy dependence of tunneling events is related to an energy-dependent phase difference between interference paths. Notice that $\langle I_c \rangle$ does not depend on the parameter $\alpha$, and this is due to the different tunneling processes which generate the two currents ($p$- or $f$- tunneling).\\
\begin{figure}[ht!]
	\centering
	\includegraphics[width=.85\linewidth]{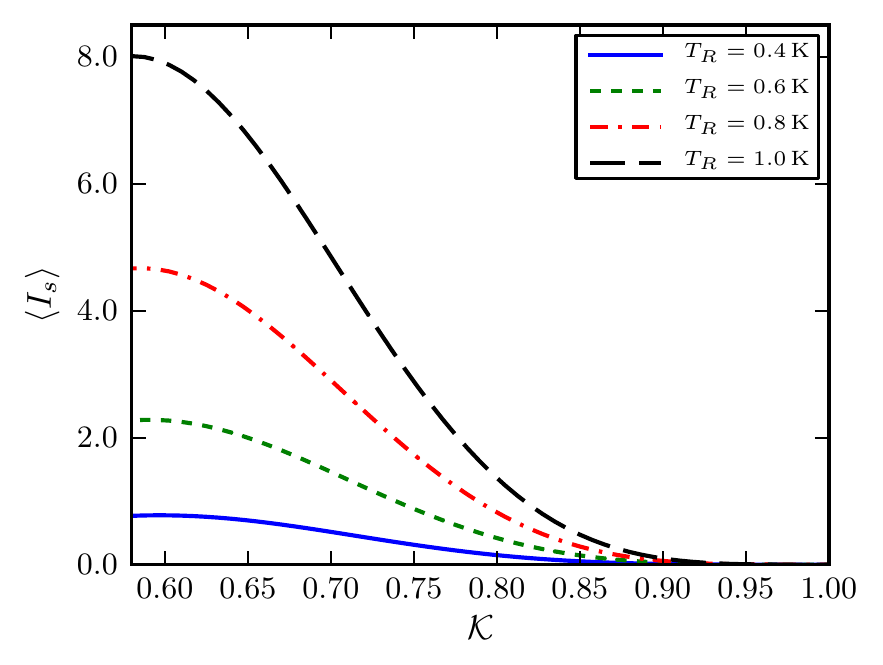}
	\caption{(Color online) Average spin current $\langle I_s \rangle$ in units of $10^{-6} |\lambda_f|^2/(8\pi)$ as a function of the parameter $\mathcal K$. Different curves correspond to the different values $T_R=0.4$ K, $0.6$ K, $0.8$ K and $1.0$ K, while $T_L$ is fixed at $T_L=0.1$ K. Parameters $\eta$ and $\alpha$ are set to $\eta=8\pi$ and $\alpha=0.4$. The chemical potential and the high energy cut-off are $\mu=100$ K and $\omega_c=200$ K.}
	\label{fig:spin_K}
\end{figure}\begin{figure}[h]
\centering
\begin{overpic}[width=0.85\linewidth]{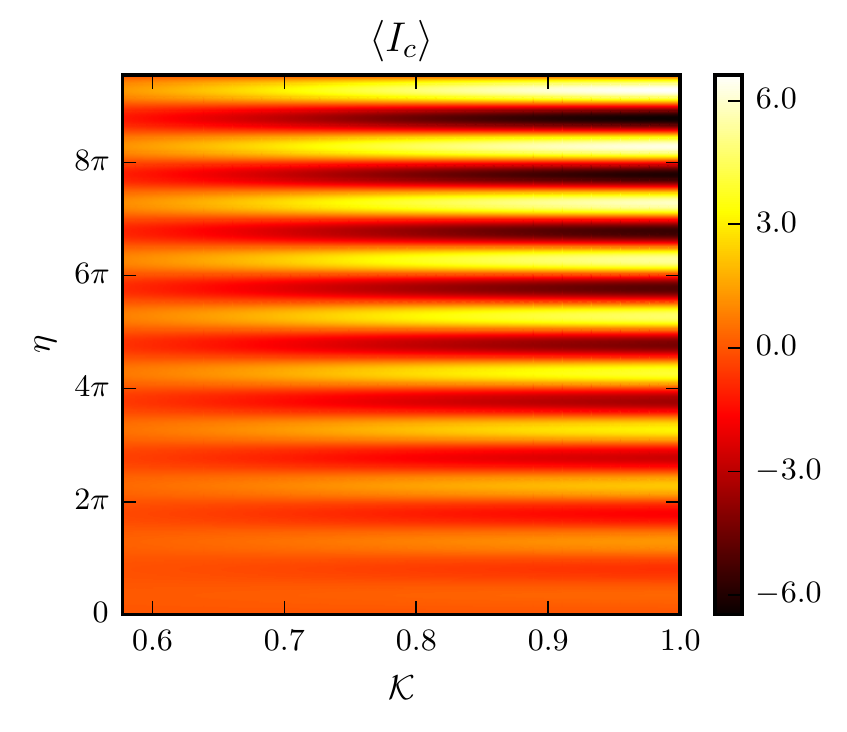}\put(1,70){a)}\end{overpic}
\begin{overpic}[width=0.85\linewidth]{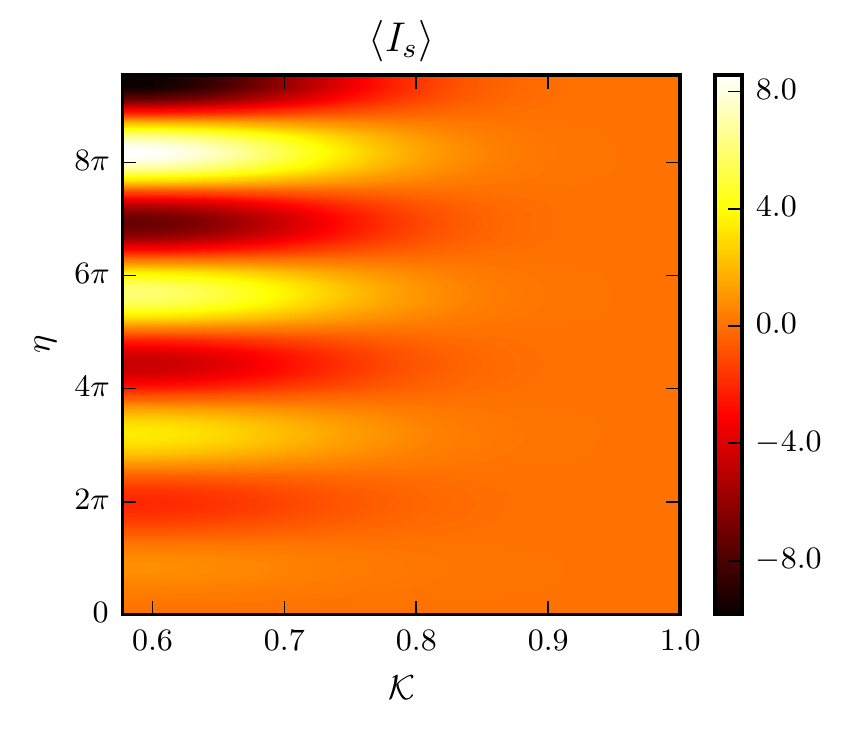}\put(1,70){b)}\end{overpic}
\caption{\small{(Color online) Density plot of charge (a) and spin (b) current $\< I_c \>$ and $\< I_s \>$ as a function of $\eta$ ($y$ axis) and $\mathcal{K}$ ($x$ axis) for a double QPC geometry, in units of $10^{-4} e|\lambda_p|^2/(2\pi)$ and $10^{-6} |\lambda_f|^2/(8\pi)$ respectively. Parameter  $\alpha$ is set to $\alpha=0.4$. The two temperatures are $T_L=0.1$ K and $T_R=1$ K. The chemical potential and the high energy cut-off are $\mu=100$ K and $\omega_c=200$ K.}}
\label{fig:density}
\end{figure}Fig.~\ref{fig:spin_K} shows the behavior of the spin current $\< I_s \>$ as a function of the interaction parameter $\cal K$ for different values of the thermal gradient. As expected, the limit ${\cal K} \to 1$ leads to a vanishing current while a finite thermoelectric spin current appears as soon as repulsive electron-electron interactions are taken into account (${\cal K}<1$). It is worth noting that the spin current increase monotonically as the thermal gradient is enhanced. For the sake of definiteness we fix henceforward the temperatures as $T_L=0.1$~K and $T_R=1$~K. We have chosen such values since (i) they are experimentally accessible \cite{Konig07,Roth09,Knez11,Li15}, (ii) a non-linear regime is required in order to observe a finite spin current and (iii) the thermal gradient is sufficiently large to get relevant signals and to increase the visibility in the generated thermoelectric currents (as shown in Fig.~\ref{fig:spin_K}). In Fig.~\ref{fig:density} the average charge current (a) and spin current (b) are shown in a density plot as a function of $\eta$ ($y$-axis) and interaction parameter ${\cal K}$ ($x$-axis). Both quantities show an oscillating behavior as a function of $\eta$ (or equivalently the distance between the two QPCs). In particular $\langle I_c \rangle$ oscillates with $\eta$ with a period governed by $\sin(2\eta)$. This sinusoidal dependence has an envelope which grows with increasing $\eta$ and eventually vanishes for very large values of $\eta$ (i.e.~for very large distance $d$).  {The zeros of charge current occur at $\eta=\frac{\pi}{2}(2n+1)$ (with $n$ integer), which correspond to the condition for resonant tunneling through the two point contacts: in this case single electron tunneling does not contribute to charge current at lowest order in tunneling as pointed out in Ref. \onlinecite{Mirlin13} and \onlinecite{Furusaki93}, and higher order tunneling processes should be considered\cite{Furusaki93}.} As a function of ${\cal K}$ the charge current has a decreasing amplitude while increasing the interaction strength (i.e.~decreasing the parameter ${\cal K}$). On the other hand, the spin current oscillates with a different period due to the presence of the parameter $\alpha$ in Eq.~\eqref{I_spin_2qpc}. In the interval $-1\le\alpha \le 1$ ($\alpha =0.4$ in the figure) less zeros are expected for the spin current with respect to the charge one. The contrary occurs for $|\alpha |>1$ and the visibility of spin current decreases with respect to charge current. For this reason, we focused the analysis on the interval $-1\le\alpha \le 1$. For each value of $\eta$ the spin current shows the qualitative behavior described in Fig.~\ref{fig:spin_K}, with $\< I_s \>$ positive or negative depending on the specific value of $\eta$. By modulating the two gate voltages (parameters $\eta$ and $\alpha$), the oscillating behavior of currents allow to induce a {\it pure} thermoelectric spin current with vanishing charge current. This is shown in Fig.~\ref{fig:ratio} with the ratio $\langle I_s \rangle/\langle I_c \rangle$ as a function of $\eta$ for three values of $\alpha$.
Configurations with a pure spin current and a vanishingly small charge current are achieved whenever the ratio diverges.\\
This fact is true also for different values of temperatures and interaction parameter.
As an example, for $\eta=2\pi$, the ratio diverges for $\alpha=0.10$ and $\alpha=0.40$. Note however that it remains finite for $\alpha=0.25$, because both charge and spin current vanish in this configuration.\\
Once the parameter $\eta$ is fixed, e.g.~in order to have a {\it pure} spin current one can still manipulate independently the parameter $\alpha$, controlling the amplitude and the period of $\< I_s \>$. This is shown in Fig.~\ref{fig:alpha}, where a pure spin current is achieved for different values of $\eta$. 

\section{Conclusions}
\label{sec:conclusions}

We have investigated thermoelectric transport properties of a 2D TI in a two terminal configuration. In particular we have calculated the charge and spin current induced by a thermal gradient in presence of a tunneling region. We have found that electron-electron interactions play a fundamental role in the generation of finite thermoelectric effect and spin current. We have considered both spin-preserving and spin-flipping tunneling processes, showing that the latter are responsible for the finite average spin current in the  interacting case. Focusing on the double quantum point contact geometry, we have demonstrated that it is also possible to generate a pure spin current acting on gate voltages or varying the distance between the two QPCs.

\section*{Acknowledgments}

{We thank A. Braggio, R. S\'anchez and B. Sothmann for useful discussions}.
We acknowledge the support of the MIUR-FIRB2012 - Project HybridNanoDev (Grant  No.RBFR1236VV), EU FP7/2007-2013 under REA grant agreement no 630925 -- COHEAT, MIUR-FIRB2013 -- Project Coca (Grant No.~RBFR1379UX) and the COST Action MP1209.
G.D. thanks also CNR-SPIN for support via Seed Project PGESE003 and the National Research Fund, Luxembourg under grant ATTRACT 7556175.


\begin{thebibliography}{9}
\bibitem{footnote} The Fermi distribution is here defined having zero energy at the chemical potential.
\end{thebibliography}
\end{document}